# BOINC: A Platform for Volunteer Computing

David P. Anderson
University of California, Berkeley
Space Sciences Laboratory
Berkeley, CA 94720

davea@berkeley.edu

December 9, 2018

**Abstract**

"Volunteer computing" is the use of consumer digital devices for high-throughput scientific computing. It can provide large computing capacity at low cost, but presents challenges due to device heterogeneity, unreliability, and churn. BOINC, a widely-used open-source middleware system for volunteer computing, addresses these challenges. We describe its features, architecture, and implementation.

# 1. Introduction

## 1.1 Volunteer computing

**Volunteer computing** (VC) is the use of consumer digital devices, such as desktop and laptop computers, tablets, and smartphones, for high-throughput scientific computing. Device owners participate in VC by installing a program that downloads and executes jobs from servers operated by science projects. There are currently about 30 VC projects in many scientific areas and at many institutions. The research enabled by VC has resulted in numerous papers in Nature, Science, PNAS, Physical Review, Proteins, PloS Biology, Bioinformatics, J. of Mol. Biol., J. Chem. Phys, and other top journals [1].

About 700,000 devices are actively participating in VC projects. These devices have about 4 million CPU cores and 560,000 GPUs, and collectively provide an average throughput of 93 PetaFLOPS. The devices are primarily modern, high-end computers: they average 16.5 CPU GigaFLOPS and 11.4 GB of RAM, and most have a GPU capable of general-purpose computing using OpenCL or CUDA.

The potential capacity of VC is much larger: there are more than 2 billion consumer desktop and laptop computers [2]. Current models have a peak performance (including GPU) of over 3 TeraFLOPS, giving a total peak performance of 6,000 ExaFLOPS. The capacity of mobile devices will soon be similar: there are about 10 billion mobile devices, and current mobile processors have on-chip GPUs providing as much as 4 TeraFLOPS [3]. Studies suggest that of people who learn about VC, between 5% and 10% participate, and that desktop and mobile devices are available to compute for VC about 60% and 40% of the time respectively [4] [5]. Taking these factors into account, the near-term potential capacity of VC is on the order of hundreds of ExaFLOPS.

The monetary cost of VC is divided between volunteers and scientists. Volunteers pay for buying and maintaining computing devices, for the electricity to power these devices, and for Internet connectivity. Scientists pay for a server and for the system administrators, programmers and web developers needed to operate the VC project. Operating a typical VC project involves a few Linux server computers and a part-time system administrator, costing the research project on the order of $100K per year. Several BOINC projects (Einstein@Home, Rosetta@home, SETI@home) are of this scale, and they average about 1 PetaFLOPS throughput each.

How much would this computing power cost on Amazon EC2? According to Ostermann et al. [6], the best computing value is a node type that costs $0.24 per hour and provides 50 GigaFLOPS. It would take 20,000 such instances to provide 1 PetaFLOPS. This would cost $42 million per year – 420 times the cost of using VC. Kondo et al. compared the cost of volunteer and cloud computing in more detail, and reached a similar conclusion [7].

In terms of power consumption, data-center nodes are more efficient (i.e. have greater FLOPS/watt) than many consumer devices. However, to compare net energy usage, two other factors must be considered. First, data centers use air conditioning to remove the heat generated by hardware; consumer devices generally do not. Second, in cool climates, the heat generated by consumer devices contributes to ambient heating, so the net energy cost of computing may be zero. Thus it's not clear whether VC is globally more or less energy efficient than data-center computing. In any case, this does not affect VC's cost to the scientist, since the energy is paid for by volunteers.

VC is best suited to **high-throughput computing** (HTC): workloads consisting of large groups or streams of jobs where the goal is high rate of job completion, rather than low job turnaround time. VC is less suited to workloads that have extreme memory or storage requirements, or for which the ratio of network communication (i.e. input and output file size) to computing is extremely high.

VC differs from other forms of HTC, such as grids and clouds, in several ways:

- The computers are anonymous and untrusted.
- The computers are heterogeneous in many hardware and software dimensions; for example, of desktops and laptops participating in VC, 85% are Windows, 7% are Macs, and 7% are Linux.
- Job turnaround times are uncertain because of sporadic host availability.
- There is significant "device churn", i.e. arrival and disappearance of computers.

- Creating the resource pool requires volunteer recruitment and retention, and the client software must be easy for consumers to install and run.
- The scale of VC is larger: up to millions of computers and millions of jobs per day.

Each of these factors presents challenges that a VC platform must address.

### 1.2 BOINC

Most VC projects use BOINC, an open-source middleware system for VC [8]. BOINC lets scientists create and operate VC projects, and lets volunteers participate in these projects. BOINC addresses the VC-specific issues listed above.

Volunteers install an application (the BOINC client) and then choose one or more projects to support. The client is available for desktop platforms (Windows, Mac, Linux) and for mobile devices running Android. BOINC is designed to compute invisibly to the volunteer. It runs jobs at the lowest process priority and limits their total memory footprint to prevent excessive paging. On mobile devices it runs jobs only when the device is plugged in and fully charged, and it communicates only over WiFi.

BOINC can handle essentially all HTC applications. Many BOINC projects run standard scientific programs such as Autodock, Gromacs, Rosetta, LAMMPS, and BLAST. BOINC supports applications that use GPUs (using CUDA and OpenCL), that use multiple CPUs (via OpenMP or OpenCL), and that run in virtual machines or Docker containers.

The development of BOINC began in 2002, and was carried out by a team at UC Berkeley led by the author, with funding from the National Science Foundation. The first BOINC-based project was launched in 2004. BOINC is distributed under the open-source LGPL v3 license, but applications need not be open-source.

## 2. The high-level structure of BOINC

BOINC's architecture involves multiple components interacting through HTTP-based RPC interfaces. It is designed to be modular and extensible.

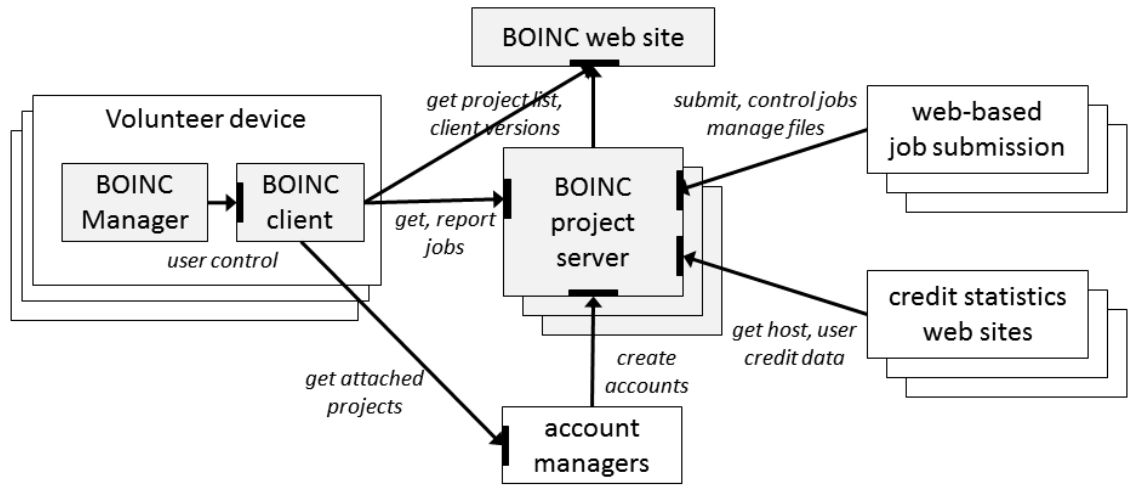

**Figure 1: The components and RPC interfaces of the BOINC system.**

Fig. 1 shows the components of BOINC; these components are described subsequently. The shaded boxes represent the BOINC software distribution; unshaded boxes represent components developed by third parties.

## 2.1 Projects and volunteers

In BOINC terminology, a **project** is an entity that uses BOINC to do computing. Each project has a **server**, based on the BOINC server software, that distributes jobs to volunteer devices. Projects are autonomous and independent. A project may serve a single research group (SETI@home), a set of research groups using a common set of applications (Climateprediction.net, nanoHUB@home) or a set of unrelated scientists (IBM World Community Grid, BOINC@TACC).

Projects may operate a public-facing web site. The BOINC server software provides database-driven web scripts that support volunteer login, message boards, profiles, leader boards, badges, and other community and incentive functions. The project may extend this with their own web content, such as descriptions and news of their research. A project is identified by the URL of its web site.

A **volunteer** is the owner of a computing device (desktop, laptop, tablet, or smartphone) who wants to participate in VC. They do so by a) installing the BOINC client on the device; b) selecting projects to support, and creating an **account** on each project, and c) **attaching** the BOINC client to each account. Each attachment can be assigned a **resource share** indicating the relative amount of computing to be done, over the long term, for the project.

## 2.2 Client/server interaction

When attached to a project, the BOINC client periodically issues RPCs to the project's server to report completed jobs and get new jobs. The client then downloads application and input files, executes jobs, and uploads output files. All communication uses client-initiated HTTP; thus the BOINC client works behind firewalls that allow outgoing HTTP, and with HTTP proxies. The server must be publicly accessible via HTTP.

Downloaded files can be compressed in transit; the client uncompresses them. Files need not be located on project servers. The validity of downloaded files is checked using hashes, and (for application files) code signing; see section 3.10.

Files are uploaded by an HTTP POST operation, which is handled by a CGI program on the server. Upload filenames include a random string to prevent spoofing. To prevent DoS attacks, projects can optionally use encryption-based tokens to limit the size of uploaded files. File uploads need not go to the project server; for example, Climateprediction.net collaborates with scientists in different countries, and the result files are sent directly to the collaborators' servers.

The client also interacts occasionally with a server operated by the BOINC project itself, to obtain current lists of projects and of client versions.

All client/server interactions handle failure using exponential back-off in order to limit the rate of requests when a server resumes after a period of being off-line.

## 2.3 Account managers

The method described above for finding and attaching projects can be cumbersome when a) the volunteer has many computers and wants to attach or detach projects on all of them, or b) the volunteer wants to attach many projects.

To address this, BOINC provides a framework for web-based **account managers** (AMs). The BOINC client, rather than being explicitly attached to projects, can be attached to an AM. The client periodically does an RPC to the AM. The reply includes a list of projects (and corresponding accounts) to which the client should attach. The client then attaches to these projects and detaches from others.

The AM architecture was designed to support "project selection" web sites, which show a list of projects with checkboxes for attaching to them. This solves the two problems described above; for example, a volunteer with many computers can attach them all to the AM account, and can then change project attachments across all the computers with a single AM interaction. Two such account managers have been developed: GridRepublic (https://www.gridrepublic.org/) and BAM! (https://boincstats.com/en/bam/). More recently, the AM architecture has been used to implement the **coordinated VC model** (see Section 10.1), in which volunteers choose science areas rather than projects.

When a volunteer selects a project on an AM, they initially have no account on that project. The AM needs to be able to create an account. To support this, projects export a set of RPCs for creating and querying accounts.

## 2.4 Computing and keyword preferences

Volunteers can specify **computing preferences** that determine how their computing resources are used. For example, they can enable **CPU throttling**, in which computing is turned on and off with a configurable duty cycle, in order to limit CPU heat.  Other options include whether to compute when the computer is in use; limits on disk and memory usage; limits on network traffic; time-of-day limits for computing and network, how much work to buffer, and so on. Computing preferences can be set through a web interface on either a project or account manager; they propagate to all the volunteer's computers that are attached to that account.  Preferences can also be set locally on a computer.

More recently, we have introduced the concept of **keyword preferences**.  BOINC defines two hierarchies of "keywords": one set for science areas (Physics, Astrophysics, Biomedicine, cancer research, etc.) and another set for location of the research project (Asia, United States, U.C. Berkeley, etc.)  Volunteers can mark keywords as "yes" or "no", indicating either that they prefer to do computing tagged with the keyword, or that they are unwilling to do it. This system has been used for two purposes:

- Projects that do research in multiple science areas can let their volunteers choose what areas to support.  The project assigns keywords to jobs.  In selecting jobs to send to a given volunteer, BOINC will prefer jobs with "yes" keywords, and won't choose jobs with "no" keywords.
- It enables the coordinated VC model. The "coordinator" account manager dynamically assigns projects to volunteers in a way that reflects project keywords and volunteer keyword preferences (see Section 10.1).

## 3. Job processing abstractions and features

### 3.1 Handling resource heterogeneity

The pool of volunteered computers is highly diverse.  At a hardware level, computers may have Intel-compatible, Alpha, MIPS, or ARM CPUs, and their 32- and 64-bit variants.  Intel-compatible CPUs can have various special instructions like SSE3. ARM processors may have NEON or VFP floating-point units. The computers run various operating systems (Windows, Mac OS X, Linux, FreeBSD, Android) and various versions of these systems.  There are a range of GPUs made by NVIDIA, AMD, and Intel, with various driver versions offering different features.

BOINC provides a framework that lets projects use as many of these computing resources as possible, and to use them as efficiently as possible.   Projects can build versions of their programs for specific combinations of hardware and software capabilities.  These are called **app versions**, and the collection of them for a particular program is called an **app**.  Jobs are submitted to apps, not app versions.

BOINC defines a set of **platforms**, which are generally the combination of a processor type and operating system: for example, (Windows, Intelx64).  Each app version is associated with a platform, and is sent only to computers that support that platform.

For finer-grained control of version selection, BOINC provides the **plan class** mechanism.  A plan class is a function that takes as input a description of a computer (hardware and software), and returns a) whether an app version can run on that computer; b) if so what resources (CPU and GPU, possibly fractional) it will use, and c) the peak FLOPS of those resources.  For example, a plan class could specify that an app version requires a particular set of GPU models, with a minimum driver version.  Plan classes can be specified either in XML or as C++ functions.  App versions can optionally be associated with a plan class.

App versions have integer version numbers.  In general, new jobs are processed using the latest version for a given (platform, plan class) pair; in-progress jobs may be using older versions.

When the BOINC client requests jobs from a server, it includes a list of platforms it supports, as well as a description of its hardware and software.  When the server dispatches a job to a computer, it selects an app version with a matching platform, and whose plan class function accepts the computer.  The details of this scheduling policy are described in Section 6.4.

As an example, SETI@home has 9 supported platforms, 2 apps, 86 app versions, and 34 plan classes.

### 3.2  The anonymous platform mechanism

By default, the BOINC client gets application executables from the project server. This model doesn't address the following scenarios:

- The volunteer's computer type is not supported by the project.
- The volunteer wants, for security reasons, to run only executables they have compiled from source.
- The volunteer wants to optimize applications for particular CPU models, or to make versions for GPUs or other coprocessors.

To handle these cases, BOINC offers a mechanism called **anonymous platform**. This lets volunteers build applications themselves, or obtain them from a third party, rather than getting them from the project server. This mechanism can be used only for projects that make their application source code available.

After building their own version of project's applications, the volunteer creates a configuration file describing these versions: their files, their CPU and GPU usage, and their expected FLOPS.  If the BOINC client finds such a configuration file, it conveys it to the scheduler, which uses that set of app versions in the job dispatch process.  Projects that support anonymous platform typically supply input and output files for a set of "reference jobs" so that volunteers can check the correctness of their versions.

In some projects (for example, SETI@home) the anonymous platform mechanism has enabled a community of volunteers who have made versions for various GPUs and vendor-specific CPU features, and who have made improvements in the base applications.  Many of these versions have later been adopted by the project itself.

### 3.3  Jobs properties

A BOINC **job** includes a reference to an app and a collection of input files.  A job has one or more **instances**.  Jobs have input files; instances have output files. When the server dispatches an instance to a client, it specifies a particular app version to use. The client downloads the input files and the files that make up the app version, executes the main program, and (if it completes successfully) uploads the resulting output files and the standard error output.

In the following we use "FLOPs" (lower-case s) as the plural of FLOP (floating-point operation). FLOPS (upper-case s) denotes FLOPs per second.

Each job has a number of properties supplied by the submitter. These include:

- An estimate of the number of FLOPs performed by the job (used to estimate runtime; see section 6.3).
- A maximum number of FLOPs, used to abort jobs that go into an infinite loop.
- An estimate of RAM working-set size, used in server job selection (see section 6.4)
- An upper bound on disk usage, used both for server job selection and to abort buggy jobs that use unbounded disk space.
- An optional set of keywords describing the job's science area.

### 3.4  Result validation

Hosts may return incorrect results because of hardware errors or malicious user behavior. Projects typically require that the final results of the job are correct with high probability. There may be application-specific ways of doing this: for example, for physical simulations one could check conservation of energy and the stability of the final configuration. For other cases, BOINC provides a mechanism called **replication-based validation** in which each job is processed on two unrelated computers. If the outputs agree, they are accepted as correct; otherwise a third instance is created and run. This is repeated until either a quorum of consistent instances is achieved, or a limit on the number of instances is reached.

What does it mean for two results to agree? Because of differences in floating-point hardware and math libraries, two computers may return different but equally valid results. McIntosh et al. [9] showed that by carefully selecting compilers, compiler options, and libraries it's possible to get bitwise-identical results for FORTRAN programs across the major platforms. In general, however, discrepancies will exist. For applications that are numerically stable, these discrepancies lead to small differences in the final results. BOINC allows projects to supply application-specific **validator** functions that decide if two results are equivalent, i.e. whether corresponding values agree within specified tolerances.

Physical simulations are typically not stable. For such applications, BOINC provides a mechanism called **homogeneous redundancy** [10] in which computers are divided into equivalence classes, based on their hardware and software, such that computers in the same class compute identical results for the app. Once a job instance has been dispatched to a computer, further instances are dispatched only to computers in the same equivalence class. BOINC supplies two equivalence relations: a coarse one involving only OS and CPU vendor, and a finer-grained one involving CPU model as well. Projects can define their own equivalence relations if needed.

In some cases there are discrepancies between app versions; for example, CPU and GPU versions may compute valid but incomparable results. To address this, BOINC provides an option called **homogeneous app version**. Once an instance has been dispatched using a particular app version, further instances use only the same app version. This can be used in combination with homogeneous redundancy.

Basic replication-based validation reduces effective computing capacity by a factor of at least two. BOINC provides a refinement called **adaptive replication** that moves this factor close to one. The idea is to identify hosts that consistently compute correct results (typically the vast majority of hosts) and use replication only occasionally for jobs sent to these hosts. Actually, since some computers are reliable for CPU jobs but unreliable for GPU jobs, we maintain this "reputation" at the granularity of (host, app version).

Adaptive replication works as follow: the BOINC server maintains, for each (host, app version) pair, a count $N$ of the number of consecutive jobs that were validated by replication. Once $N$ exceeds a threshold, jobs sent to that host with that app version are replicated only some of the time; the probability of replication goes to zero as $N$ increases. Adaptive replication can achieve a low bound on the error rate (incorrect results accepted as correct), even in the presence of malicious volunteers, while imposing only a small throughput overhead.

Result validation is useful for preventing credit cheating (see section 7). Credit is granted only to validated results, and "faked" results generally will fail validation. For applications that usually return the same answer (e.g. primality checks), cheaters could simply always return this result, and would usually get credit. To prevent this, the application can add extra output that depends on intermediate results of the calculations.

### 3.5 Specialized job processing features

App versions are assigned version numbers that increase over time. In general, BOINC always uses the latest version for a given (platform, plan class) combination. In some cases a project may release a new app version whose output is different from, and won't validate against, the previous version. To handle this, BOINC allows jobs to be "pinned" to a particular version number, in which case they will be processed only by app versions with that version number.

Some BOINC projects have jobs that run for weeks or months on typical computers. Because of host churn or user impatience, some of these jobs don't run to completion. The intermediate outputs of these jobs may have scientific value. BOINC has two features to support such jobs. First, applications can tell the client that a particular output file is complete and can be uploaded immediately, rather than waiting for the job to finish. Second, the application can generate **trickle-up messages** that are conveyed immediately to the server and handled by project-specific logic. This can be used, for example, to give volunteers credit for partial job completion.

Some applications have large input files, each of which is used by many jobs. To minimize network traffic and server load for these applications, BOINC has a feature called **locality scheduling**. The large input files are marked as **sticky** (see Section 3.10) so that they remain resident on the host. The scheduler RPC request message includes a list of sticky files on the host. The scheduler preferentially sends jobs that use these files (see Section 3.5).

BOINC allows jobs to be **targeted** at a particular host or volunteer. This can be used, for example to ensure that test jobs are executed on the project's own computers.

The difference in performance between volunteer devices – say, a phone versus a desktop with a high-end GPU – can be several orders of magnitude. A job that completes in five minutes on one could take a week on the other. To reduce server load and maintain volunteer interest, it's desirable that job runtime be on the order of an hour. To support uniformity of job runtime across platforms, BOINC provides a mechanism that lets projects create jobs in a range of sizes (say, processing smaller or larger amounts of data). The BOINC server maintains statistics of the performance of each (host, app version) pair and dispatches jobs to appropriate hosts based on their size – larger jobs are sent to faster hosts.

Applications that use negligible CPU time (e.g., sensor monitoring, network performance testing, and web crawling) can be labeled as **non-CPU-intensive**. BOINC treats these applications specially: it runs a single job per client, it always runs the application, and it runs it at normal process priority.

Debugging BOINC applications – e.g. figuring out why they crash – is challenging because the errors occur on remote, inaccessible machines. BOINC provides features to assist in this. The client reports exit codes and signal numbers, as well as standard error output, to the server. The server provides web interfaces allowing developers to break down errors by platform, app version, and error type, and to see the details of devices where errors occurred; this facilitates finding problems specific to a particular OS version or video device driver.

### 3.6 The BOINC Runtime Environment

We now discuss the environment in which the BOINC client runs project applications, and how the client controls and communicates with running jobs.

The BOINC client installers for Windows and Mac OS X create an unprivileged user account under which BOINC runs applications. This provides a measure of protection against buggy or malicious applications, since they are unable to read or write files belonging to existing users.

The BOINC client stores its files in a dedicated directory. Within this directory, there is a **project directory** for each project to which the client is attached, containing the files associated with that project. There are also a number of **job directories**, one for each in-progress job. Job directories contain symbolic links to files in the corresponding project directory, allowing multiple concurrent jobs to share single copies of executable and input files.

To enforce volunteer preferences, to support CPU throttling, and to support the client GUI's monitoring and control functions, the client must be able to suspend, resume, and abort jobs, and to monitor their CPU time and memory usage. Jobs may consist of a number of processes. Some modern operating systems provide the needed process-group operations, but not all do, and BOINC must run on older systems as well.  So BOINC implements these functions itself, based on message-passing between the client and the application.  This message-passing uses shared memory, which is supported by all operating systems.  There are queues for messages from client to application (suspend, resume, quit) and from application to client (current CPU time, CPU time at last checkpoint, fraction done, working set size).  Each side polls its incoming queues every second.

All applications run under BOINC must implement this message protocol.  One way to do this is to rebuild the application for the target platforms, linking it with a BOINC runtime library, and using its API; we call this a **native** application.  Another way is to use an existing executable together with a **wrapper** program, as described in the next section.  These two approaches are shown in Fig. 2.

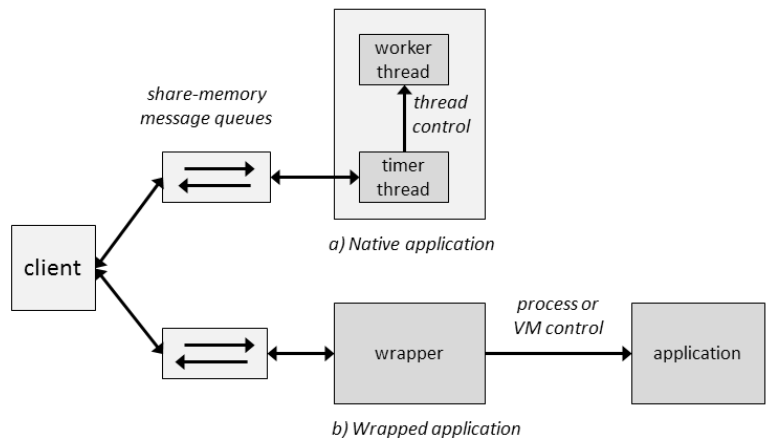

**Figure 2: The process structure of the BOINC runtime environment.**

For sequential programs, the runtime library creates a **timer thread** that manages communication with the client, while the original **worker thread** runs the program.  On Windows, the timer thread controls the worker thread with Windows thread primitives.  The Unix thread library doesn't let one thread suspend another, so we use a 10 Hz signal, handled by the worker thread, and suspend the worker by having this signal handler sleep.

For multithread applications (e.g., OpenMP or OpenCL CPU) the process structure is a bit different. On Unix, the program forks.  The parent process handles process control messages, which it implements using signals.  The child process runs the program, and uses a timer thread to handle status and checkpointing messages.

BOINC supports application-level checkpoint/restart.  Every few minutes the client sends a message asking the application to checkpoint.  When the application reaches a point where it can efficiently checkpoint (e.g. the start of its outer loop) it writes a checkpoint file and sends a message to the client indicating this.  Thus the client knows when applications have checkpointed, and can avoid preempting jobs that haven't checkpointed in a long time, or ever.

In applications that use GPUs, the CPU part of the program dispatches short "kernels" to the GPU.  It's imperative that the application not be suspended while a kernel is in progress.  So the BOINC runtime library supports **masked sections** during which suspension and abortion are deferred. Execution of GPU kernels (as well as writing checkpoint files) should be done in a masked section.

By default, the BOINC client runs applications at the lowest process priority. Non-CPU-intensive and wrapper applications are run at normal priority. GPU applications have been found to perform poorly if run at low priority, so they're also run at normal priority.

CPU throttling (section 2.4) is implemented by having the client suspend and resume the application, with one-second granularity, and with a duty cycle determined by the volunteer's setting.

Applications may encounter a transient situation where they can't continue; for example, a GPU application may fail to allocate GPU memory. The BOINC API provides a **temporary exit** function that causes the application to exit and tells the client to schedule it again after a certain amount of time. Jobs that do this too many times are aborted.

Applications should make minimal assumptions about host software. If they need libraries other than standard system libraries, they must include them. When the BOINC client runs a job, it modifies its library search path to put the job directory first, followed by the project directory.

### 3.7  Application graphics

App versions can include a second executable program that generates a visualization of the computation in progress. The graphics application may be run by the BOINC screensaver (which cycles among the graphics of all running jobs) or by a command from the BOINC Manager.

The graphics application for a job is run in the job's directory. It may get state information by reading files in that directory, or by creating a shared memory segment shared with the application. Graphics programs typically use OpenGL for portability.

### 3.8  Wrapped and Virtual Machine applications

Native applications must be modified to use the BOINC API, and then rebuilt for the target platforms. This may be difficult or impossible, so BOINC provides "wrapper" mechanisms for using existing executables.

The **BOINC wrapper** allows arbitrary executables to be run under BOINC, on the platforms for which they were built. The wrapper acts as an intermediary between the BOINC client and the application. It handles runtime messages from the client (section 3.6) and translates them into OS-specific actions (e.g. signals on Unix) to control the application. BOINC provides versions of the wrapper for the major platforms. An app version includes the appropriate version of the wrapper (as the main program), and the application executable. The wrapper can be instructed to run a sequence of executables. Marosi et al. developed a more general wrapper that supports arbitrary workflows using a shell-like scripting language [11].

BOINC supports applications that run in virtual machines using VirtualBox, a multi-platform open-source VM system for Intel-compatible computers. Applications can then be developed on Linux and run on Windows and Mac OS X computers with no additional work. The BOINC client detects and reports the presence of VirtualBox. VM apps are only dispatched to hosts where VirtualBox is installed. On Windows, the recommended BOINC installer installs VirtualBox as well.

To support VM apps, BOINC provides a **VBox wrapper** program, which interfaces between the BOINC client and the VirtualBox executive. It translates runtime system messages to VirtualBox operations. VirtualBox supports sharing of directories between host and guest systems. The VBox wrapper creates a shared directory for VM jobs, places input files there, and retrieves output files from there.

In addition to reducing heterogeneity problems, VM technology provides a strong security sandbox, allowing untrusted applications to be used. It also provides an application-independent checkpoint/restart capability, based on the VirtualBox "snapshot" feature. The wrapper tells VirtualBox to create a snapshot every few minutes. If computing is stopped (e.g. because the host is turned off) the job can later be restarted, on that host, from the most recent snapshot.

BOINC's VM system can be used together with container systems like Docker [12]. Docker-based applications consist of a Docker image (the OS and library environment of choice) and an executable to run in the container. The container is run within a VirtualBox virtual machine. Docker images consist of a layered set of files. Typically only the top layer changes when new app versions are released, so network traffic and storage requirements are smaller than when using monolithic VM images.

VM apps normally include the VM or Docker image as part of the app version, but it's possible to have it be an input file of the job, in which case a single (BOINC) app can be used for arbitrarily many science applications. This is significant because creating an app includes signing its files, which is a manual operation. We call such an app a **universal app**. If a project's science applications are already packaged as VM or Docker images, they can be run under BOINC with no porting or other work.

There is a security consideration in the use of universal apps, since job input files are not code-signed (see Section 3.10). If a project's server is hacked, the hackers could substitute a different VM image. However, the danger is small since the VM is sandboxed and can't access files on the host other than its own input and output files.

VM apps were proposed in 2007 by González et al. [13], and VM wrappers were developed at CERN [14] and SZTAKI [15]. Ferreira et al. describe a wrapper able to interface to multiple VM hypervisors [16].

### 3.9 Job submission and file management

Job submission consists of three steps: 1) **staging** the job's input files to a publicly-accessible web server, 2) creating the job, and 3) handling the results of a completed job. BOINC supplies C++ APIs and command-line tools for staging files and submitting jobs on the server. It also provides HTTP RPC APIs for remote file management and job submission, and it supplies bindings of these interfaces in Python,

PHP, and C++.  These interfaces are defined in terms of "batches" of jobs, and are designed for efficiency: for example, submitting a batch of a thousand jobs takes less than a second.

The handling of completed jobs is done by a per-app **assimilator** daemon.  This is linked with a project-supplied C++ or Python function that handles a completed job.  This function might, for example, move the output files to a different directory, or parse the output files and write the results to a database.

These mechanisms make it straightforward to use BOINC as a back end for other job processing systems.  For example, we implemented a system where HTCondor jobs can be automatically routed to a BOINC project.  The mechanisms can also be used for web-based remote job submission systems.  For example, a science gateway might automatically route jobs, submitted via the web, to either local compute nodes or a BOINC server (see Section 10.2).

In some BOINC projects, multiple job submitters contend for the project's resources.  BOINC provides an allocation system that handles this contention in a way that's fair to submitters both with sporadic and continuous workloads.  For each submitter, the system maintains a **balance** that grows linearly at a particular rate, up to a fixed maximum.  The rate may differ between submitters; this determines the average computing throughput available to each submitter. When a submitter uses resources, their balance is decreased accordingly.  At any given point, the jobs of the submitter with the greatest balance are given priority.  We call this the **linear-bounded model**; BOINC uses it in other contexts as well (sections 6.1 and 10.1).  Given a mix of continuous and sporadic workloads, this policy prioritizes small batches, thereby minimizing average batch turnaround.

### 3.10  Storage

The BOINC storage model is based on named files. Files are immutable: a file with a given name cannot change.  Projects must enforce this.  Input and output files also have **logical names**, by which applications refer to them; these need not be unique.

The entities involved in computation – app versions and jobs – consist of collections of files.  The files in an app version must be **code-signed** by tagging them with a PKE-encrypted hash of the file contents. The recommended way of doing this is to keep the private key on a computer that is never connected to a network and is physically secure.  Files are manually copied to this computer – say, on a USB drive – and signed.  This procedure provides a critical level of security: hackers are unable to use BOINC to run malware on volunteer hosts, even if they break into project servers.

Job input and output files normally are deleted from the volunteer host after the job is finished.  They can be marked as **sticky**, in which case they are not deleted.  Scheduler requests include a list of sticky files on the client, and replies can direct the client to delete particular sticky files.

Files making up app versions are automatically sticky in the sense that they are deleted only when the app version has been superseded by another app version with the same app, platform, and plan class, and with a greater version number.

The amount of storage that BOINC may use on a given host is limited by its free disk space, and generally also by volunteer preferences (which can specify, e.g., minimum free space or maximum fraction of space used by BOINC). Scheduler requests include the amount of usable space, and the scheduler doesn't send jobs whose projected disk usage would exceed this. If the amount of usable space is negative (i.e. BOINC has exceeded its limit) the scheduler returns a list of sticky files to delete.

If the host is attached to multiple projects and the disk limit is reached, the projects are contending for space. The client computes a **storage share** for each project, based on its resource share and the disk usage of the projects; this is what is conveyed to the scheduler.

Files need not be associated with jobs. A scheduler reply can include lists of arbitrary files to be downloaded, uploaded, or deleted. Thus BOINC can be used for **volunteer storage** as well as computing. This might be used for a variety of purposes: for example, to cache recent data from a radio telescope so that, if an event is detected a few hours after its occurrence, the raw data would still be available for reanalysis. It can also be used for distributed data archival; see Section 10.3.

## 4. The life of a job

In most batch systems the life of a job is simple: it gets dispatched to a node and executed. In BOINC, however, a dispatched job has various possible outcomes:

1. The job completes successfully and correctly.
2. The program crashes.
3. The job completes but the returned results are incorrect, either because of hardware malfunction or because a malicious volunteer intentionally substitutes wrong results.
4. The job is never returned, because (for example) the computer stops running BOINC.

To ensure the eventual completion of a job, and to validate its results, it may be necessary to create additional instances of the job.

Each job *J* is assigned a parameter *delay_bound(J)* by its submitter. If an instance of *J* is dispatched to a host at time *T*, the **deadline** for its completion is *T + delay_bound(J)*. If the completed results have not been returned to the server by the deadline, the instance is assumed to be lost (or excessively late) and a new instance of *J* is created. The scheduler tries to dispatch jobs only to hosts that are likely, based on runtime estimates and existing queue sizes, to complete them by their deadlines.

Projects may have time-sensitive workloads: e.g., batches of jobs that should be finished quickly because the submitter needs the results to proceed. Such jobs may be assigned low delay bounds; however, this may limit the set of hosts to which the job can be dispatched, which may have the opposite of the desired effect.

For workloads that are purely throughput-oriented, *delay_bound(J)* can be large, allowing even slow hosts to successfully finish jobs. However, it can't be arbitrarily large: while the job is in progress, its input and output files take up space on the server, and *delay_bound(J)* limits this period. For such workloads, a delay bound of a few weeks may be appropriate.

In addition to *delay_bound(J)*, a job has the following parameters; these are assigned by the submitter, typically at the level of app rather than job:

- *min_quorum(J)*: validation is done when this many successful instances have completed. If a strict majority of these instance have equivalent results, one of them is selected as the **canonical instance**, and that instance is considered the correct result of the job.
- *init_ninstances(J)*: the number of instances to create initially. This must be at least *min_quorum(J)*; it may be more in order to achieve a quorum faster.
- *max_error_instances(J)*: if the number of failed instances exceeds this, the job as a whole fails and no more instances are created. This protects against jobs that crash the application.
- *max_success_instances(J)*: if the number of successful instances exceeds this and no canonical instance has been found, the job as a whole fails. This protects against jobs that produce nondeterministic results.

So the life of a job *J* is as follows:

- *J* is submitted, and *init_ninstances(J)* instances are created and eventually dispatched.
- When an instance's deadline is reached and it hasn't been reported, the server creates a new instance.
- When a successful instance is reported and there is already a canonical instance, the new instance is validated against it to determine whether to grant credit (see section 7). Otherwise, if there are at least *min_quorum(J)* successful instances, the validation process is done on these instances. If a canonical instance is found, the app's assimilator is invoked to handle it. If not, and the number of successful instances exceeds *max_success_instances(J)*, *J* is marked as failing.
- When a failed instance is reported, and the number of failed instances exceeds *max_error_instances(J)*, *J* is marked as failing. Otherwise a new instance is created.

When a canonical instance is found, *J*'s input files and the output files of other instances can be deleted, and any unsent instances are cancelled.

It's possible that a successful instance of *J* is reported after a canonical instance is found. It's important to grant credit to the volunteer for the instance, but only if it's correct. Hence the output files of the canonical instance are retained until all instances are resolved.

After all instances are resolved, and the job has succeeded or failed, the job and job instance records can be purged from the database. This keeps the database from growing without limit over time; it serves as a cache of jobs in progress, not an archive of all jobs. However, the purging is typically delayed for a few days so that volunteers can view their completed jobs on the web.

# 5. Software architecture

## 5.1 Server architecture

A BOINC project operates a **server** consisting of one or more computer systems running the BOINC server software. These systems may run on dedicated hardware, VMs, or cloud nodes. The server has many interacting processes, as shown in Fig. 3.

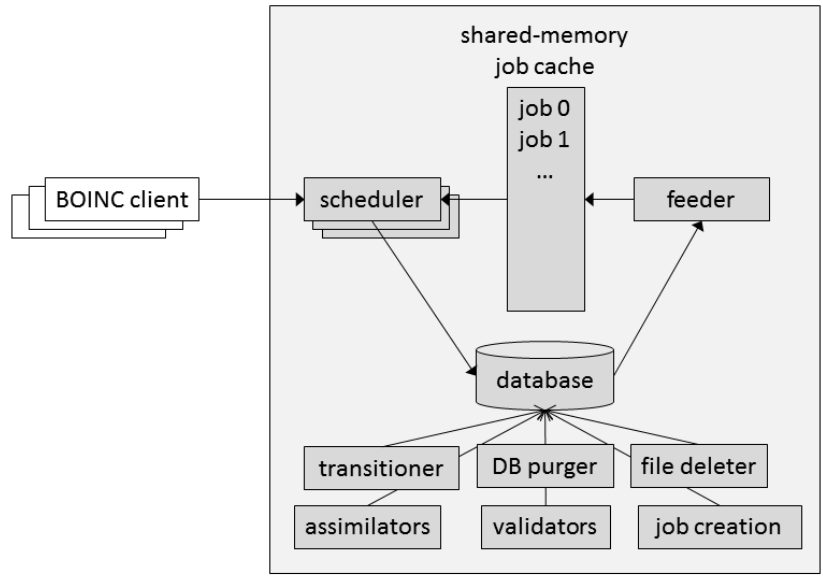

**Figure 3: The components of a BOINC server.**

A BOINC server centers on a relational database (usually MySQL or MariaDB). The database includes tables for most of the abstractions mentioned earlier: volunteers, hosts, apps, app versions, jobs, job instances, and so on. To reduce DB load, details such as the list of a job's input files are stored in XML "blobs" rather than in separate tables.

Client RPCs are handled by a **scheduler**, which is implemented as a CGI or FCGI program run from a web server. Typically many RPCs are in progress at once, so many instances of the scheduler are active. The main function of the scheduler is to dispatch job instances to clients, which requires scanning many jobs and job instances. Doing this by directy querying the database would limit performance. Instead, BOINC uses an architecture in which a shared-memory segment contains a cache of job instances available for dispatch – typically on the order of a thousand of them. This cache is replenished by a **feeder** process, which fetches unsent job instances from the database and puts them in vacant slots in the cache. The feeder supports the homogeneous redundancy, homogeneous app version, and job size features by ensuring that the shared-memory cache contains job of all the different categories.

An instance of the scheduler, rather than querying the database for jobs to send, scans the job cache. It acquires a mutex while doing this; the scan is fast so there is little contention for the mutex. When the scheduler sends a job, it clears that entry in the cache, and the entry is eventually refilled by the feeder. The scheduler updates the database to mark the job instance as sent. The efficiency of this mechanism allows a BOINC server – even on a single machine – to dispatch hundreds of jobs per second [17].

The server's other functions are divided among a set of daemon processes, which communicate and synchronize through the database.

- **Validator**: There is one for each app that uses replication. It examines the instances of a job, compares their output files, decides whether there is a quorum of equivalent results, and if so designates one instance as "canonical". For apps that use homogeneous redundancy to achieve bitwise agreement between instances, BOINC supplies a validator that compares the output files byte for byte. Other apps use custom validators with a project-supplied function that does the (fuzzy) comparison for that app.
- **Assimilator**: This handles completed and validated jobs. There is one assimilator per app; it includes a project-supplied handler function. This function might move the output files to a particular destination, or parse them and write the results to a database.
- **Transitioner**: Viewing the progress of a job as a finite-state machine, this handles the transitions. The events that trigger transitions come from potentially concurrent processes like schedulers and validators. Instead of handling the transitions, these programs set a flag in the job's database record. The transitioner enumerates these records and processes them. This eliminates the need for concurrency control of DB access.
- **File deleter**: This deletes input and output files of completed and assimilated jobs.
- **Database purger**: This deletes the database records of completed jobs and job instances.

This multi-daemon architecture provides a form of fault-tolerance. If a daemon fails (say, an assimilator fails because an external database is down) the other components continue; work for the failed component builds up in the database and is eventually processed.

A BOINC server is highly scalable. First, the daemons can be run on separate hosts. Second, all of the server functions except the database server can divided among multiple processes, running on the same or on different hosts, by partitioning the space of database job IDs: if there are $N$ instances of a daemon, instance $i$ handles rows for which *(ID mod N) = i*. In combination with the various options for scaling database servers, this allows BOINC servers to handle workload on order of millions of jobs per day.

The server software also includes scripts and web interfaces for creating projects, stopping/starting projects, and updating apps and app versions.

## 5.2 Client architecture

The BOINC client software consists of three programs, which communicate via RPCs over a TCP connection:

- The **core client** manages job execution and file transfers. The BOINC installer arranges for the client to run when the computer starts up. It exports a set of RPCs to the GUI and the screensaver.
- A GUI (the **BOINC Manager**) lets volunteers control and monitor computation. They can, for example, suspend and resume CPU and GPU computation, and control individual jobs. They can view the progress and estimated remaining time of jobs, and can view an event log showing a configurable range of internal and debugging information.
- An optional **screensaver** shows application graphics (section 3.7) for running jobs when the computer is idle.

Third party developers have implemented other GUIs; for example, BoincTasks is a GUI for Windows that can manage multiple clients [18]. For hosts without a display, BOINC supplies a command-line program providing the same functions as the Manager.

The core client is multi-threaded. The main thread uses the POSIX *select*() call to multiplex concurrent activities (network communication and time-consuming file operations like decompression and checksumming). Each GUI RPC connection is serviced by a separate thread. All the threads synchronize using a single mutex; the main thread releases this mutex while it's waiting for I/O or doing a file operation. This ensures rapid response to GUI RPCs.

## 6. Scheduling

Job scheduling is at the core of BOINC. The scheduling system consists of three interrelated parts (see Fig. 4):

- The BOINC client maintains a queue of jobs; these jobs use various combinations of CPUs and GPUs. **Resource scheduling** is the decision of what jobs to run at a given point.
- **Work fetch** involves deciding when to request more jobs from a project, how much work to request, and, if the client is attached to multiple projects, which project to ask.
- **Job selection** is the server's choice of what jobs to send in response to a client work request.

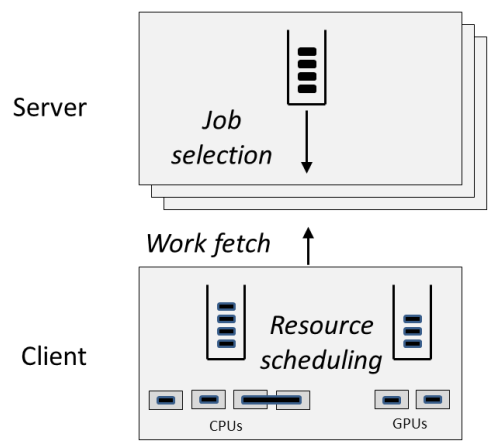

**Figure 4: Job scheduling in BOINC involves three interacting policies: resource scheduling, work fetch, and job selection.**

The scheduling system has several goals:

- Eventually complete all jobs.
- Maximize throughput by a) avoiding idle processing resources; b) completing jobs by their deadline, thereby avoiding creating additional instances of the job; c) using the best-performance app version for a given job and host.
- Enforce the volunteer's per-project resource shares and computing preferences.

BOINC's scheduling policies revolve around estimating the runtime of a job, on a particular host, using a particular app version. There are actually two notions of runtime:

- The **raw runtime** is how long the job takes running full-time on the host.
- The **scaled runtime** is how long it takes given a) CPU throttling (section 2.4) and b) the possibly sporadic availability of the processing resources used by the app version (resources are unavailable when the computer is turned off or their use is disallowed by user preferences). The client maintains averages of the availability of CPUs and GPUs, and reports these to the server.

Runtime estimates are intrinsically uncertain. BOINC tries to make good estimates, and to deal gracefully with situations where the estimates are inaccurate.

## 6.1 Client: resource scheduling

On a given host, the client manages a set of **processing resources**, consisting of a CPU and possibly one or more coprocessors such as GPUs. There may be multiple instances of each processing resource. Volunteer preferences may limit the number of available CPU instances, denoted *n_usable_cpus*. The client estimates the **peak FLOPS** of each instance. For CPUs this is the Whetstone benchmark result; for GPUs it's the vendor-supplied estimate based on number of cores and clock rate.

At a given point the client has a queue of jobs to run. A given job may be a) not started yet; b) currently executing; c) in memory but suspended, or d) preempted and not in memory.

Each job *J* has a fixed **resource usage**: the number (possibly fractional) of instances of each processing resource that it uses while executing. For CPUs, fractional resource usage refers to time: 0.5 means that *J* has 1 thread that runs half the time. For GPUs, fractional usage typically refers to core usage: 0.5 means that *J* uses half the GPU's cores. In either case, throughput is maximized by running 2 such jobs at once.

Each job *J* has an estimated RAM working set size *est_wss(J)*. If *J* is running or preempted this is the latest working set size reported by the operating system. If *J* is unstarted and there's a running job $J_2$ using the same app version, *est_wss(J)* is the working set size of $J_2$; otherwise it's the project-supplied RAM usage estimate.

We call a set of jobs **feasible** if a) for each coprocessor, the sum of usages by the jobs is at most the number of instances; b) the sum of CPU usages of CPU jobs is at most *n_usable_cpus*, and the sum of CPU usages of all jobs is at most *n_usable_cpus+1*; c) the sum of *est_wss(J)* is at most the amount of available RAM.  A set of jobs is **maximal** if it's feasible and adding any other queued job would make it infeasible.  The client always runs a maximal set of jobs.

The basic resource scheduling policy is **weighted round robin** (WRR).  Projects have dynamic **scheduling priorities** based on their resource share and recent usage, using the linear bounded model described in section 3.7.  A maximal set of jobs from the highest-priority projects are run FIFO.  At the end of each time slice (default 1 hour) priorities are recomputed and a possibly different maximal set of jobs is run.  Time-slicing is used so that volunteers see a mixture of projects.

The WRR policy can result in jobs missing their deadlines; to avoid this, we need a more sophisticated policy that predicts and avoids deadline misses. This requires estimating the remaining runtime of jobs. If a job hasn't started yet this is the job's FLOP estimate (section 3.3) divided by the server-supplied FLOPs/second estimate (section 6.4); we call this the **static estimate**.  If the job has started and has reported a fraction done (see Section 3.6) this gives a **dynamic estimate**, namely the current runtime divided by the fraction done.  For some applications (e.g. those that do a fixed number of iterations of a deterministic loop) the dynamic estimate is accurate; projects can flag such applications, instructing the client to always use the dynamic estimate.  For other applications the fraction done is approximate; in this case the client uses a weighted average of the static and dynamic estimates, weighted by the fraction done.

The client periodically simulates the execution of all jobs under the WRR policy, based on the estimated remaining scaled runtimes of the jobs.  This simulation (described in more detail below) predicts which jobs will miss their deadlines under WRR.

We can now describe BOINC's resource scheduling policy.  The following steps are done periodically, when a job exits, and when a new job becomes ready to execute:

- Do the WRR simulation.
- Make a list of jobs ordered by the following criteria (in descending priority): a) prefer jobs that miss their deadline under WRR simulation, and run these earliest deadline first; b) prefer GPU jobs to CPU jobs;  c) prefer jobs in the middle of a time slice, or that that haven't checkpointed; d) prefer jobs that use more CPUs; e) prefer jobs from projects with higher scheduling priority.
- Scan through this list, adding jobs until a maximal set is found.  Run these jobs, and preempt running jobs not in the set.

In essence, BOINC uses WRR scheduling unless there are projected deadline misses, in which case it uses earliest deadline first (EDF).  EDF is optimal on uniprocessors but not multiprocessors [19]; a policy such as least slack time first might work better in some cases.

## 6.2  Client: work fetch policy

Volunteer preferences include lower and upper limits, denoted $B_{LO}$ and $B_{HI,}$ on the amount of buffered work, as measured by remaining estimated scaled runtime.  When the amount of work for a given processing resource falls below $B_{LO}$, the client issues a request to a project, asking for enough work to bring the buffer up to at least $B_{HI}$.

This buffering scheme has two purposes. First, the client should buffer enough jobs to keep all processing resources busy during periods when no new jobs can be fetched, either because the host is disconnected from the Internet, project servers are down, or projects have no jobs. $B_{LO}$ should reflect the expected duration of these periods. Second, the frequency of scheduler RPCs should be minimized in order to limit load on project servers; an RPC should get many jobs if possible. The difference $B_{HI} - B_{LO}$ expresses this granularity.

The WRR simulation described above also serves to estimate buffer durations; see Fig. 5.

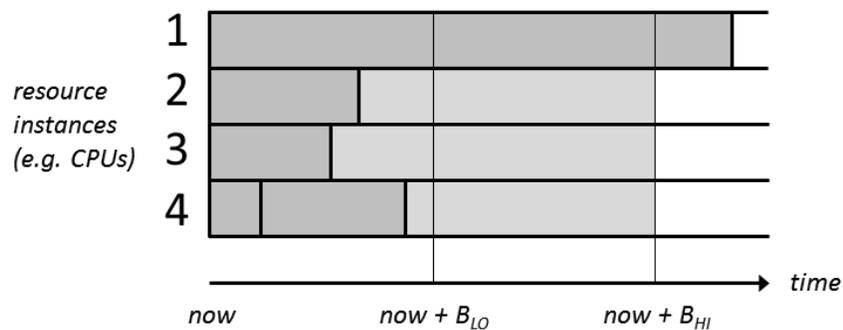

**Figure 5: Example of WRR simulation for a given processing resource. Dark gray bars represent instance usage. The area of light gray bars is the resource's "shortfall".**

In the simulation, each instance *A* of the processing resource *R* is used for some period *T(A)* starting at the current time. If $T(A) < B_{LO}$ for any instance, then the buffer for that resource needs to be replenished. *shortfall(R)* denotes the minimum additional job duration needed to use bring *T(A)* up to at least $B_{HI}$ for all instances *A*, namely

$$shortfall(R) = \sum_{instances\ A\ of\ R} \max(0, B_{HI} - T(A))$$

A work request includes several components for each processing resource *R*:

- *req_runtime(R)*: the buffer shortfall. The server should send a set of jobs whose scaled runtime estimates times the resource usage exceeds this.
- *req_idle(R)*: The number of idle instances. The server should send jobs whose total usage of *R* is at least this.
- *queue_dur(R)*: the estimated remaining scaled runtime of queued jobs that use *R*. This is used to estimate the completion time of new jobs (section 6.4).

There are situations in which a work request can't be issued to a project *P*; for example, *P* is suspended by the user, or the client is in exponential backoff from a previous failed RPC to *P*. There are also situations where work for *R* can't be requested from *P*; for example, if it has no app versions that use *R*, or user preferences prohibit it. If none of these conditions hold, we say that *R* is **fetchable** for *P*.

We can now describe the work fetch policy. First, perform the WRR simulation. If some resource needs to be replenished, scan projects in order of decreasing scheduling priority (Section 6.1). Find the first one *P* for which some processing resource *R* needs to be replenished, and *R* is fetchable for *P*. Issue a scheduler RPC to *P*, with work requests parameters as described above for all processing resources that are fetchable for *P*.

Sometimes the client makes a scheduler RPC to a project *P* for reasons other than work fetch - for example, to report a completed job instances, because of a user request, or because an RPC was requested by the project. When this occurs, the client may "piggyback" a work request onto the RPC. For each processing resource type *R*, it sees if P is the project with highest scheduling priority for which *R* is fetchable. If so, it requests work for *R* as described above.

The client doesn't generally report completed jobs immediately; rather, it defers until several jobs can be reported in one RPC, thus reducing server load. However, it always reports a completed job when its deadline is approaching, or if the project has requested that results be reported immediately.

### 6.3  Server: job runtime estimation

We now turn to the server's job selection policy. Job runtime estimation plays two key roles in this. First, the scheduler uses runtime estimates to decide how many jobs satisfy the client's request. Second, it uses the estimated turnaround time (calculated as runtime plus queue duration) to decide whether the job can be completed within its delay bound.

The job submitter may have *a priori* information about job duration; for example, processing a data file that's twice as large as another might be known to take twice as long. This information is expressed in a parameter *est_flop_count(J)* supplied by the job submitter (see section 3.3). This is an estimate of the number of FLOPs the job will use, or more precisely *TR*, where *T* is the expected runtime of the job on a processing resource whose peak FLOPs/second is *R*. If job durations are unpredictable, *est_flop_count()* can be a constant.

The BOINC server maintains, for each pair of host *H* and app version *V*, the sample mean $\overline{R}(H, V)$ and variance of *(runtime(J)/est_flop_count (J))* over all jobs *J* processed by *H* and *V*. It also maintains, for each app version *V*, the sample mean $\overline{R}(V)$ of *(runtime(J)/est_flop_count (J))* for jobs processed using *V* across all hosts.

For a given host *H* and app version *V*, the server computes a **projected FLOPS** *proj_flops(H, V)*, which can be thought of as the estimated FLOPS, adjusted for systematic error in *est_flop_count(). proj_flops(H, V)* is computed as follows.

- It the number of samples of $\overline{R}(H, V)$ exceeds a threshold (currently 10) then *proj_flops(H, V)* is $\overline{R}(H, V)$.
- Otherwise, if the number of samples of $\overline{R}(V)$ exceeds a threshold, *proj_flops(H, V)* is $\overline{R}(V)$.
- Otherwise, *proj_flops(H, V)* is the **peak FLOPS** of *V* running on *H*, i.e. the sum over processing resources *R* of *V*'s usage of *R* times the peak FLOPS of *R*.

The estimated raw runtime of job *J* on host *H* using app version *V*, denoted *est_runtime(J, H, V)*, is then *est_flop_count(J)/proj_flops(H, V)*. *proj_flops(H, V)* is also used to select which app version to use to process *J* on *H*, as explained below.

### 6.4 Server: Job dispatch policy

A scheduler RPC request message from a host *H* includes a description of *H*, including its OS, processing resources (CPUs and GPUs), memory and disk. It also includes, for each processing resource R, the fraction of time R is available for use, and the request parameters described in section 6.2

The RPC reply message includes a list of jobs, and for each job a) the app version to use, and b) an estimate of the FLOPS that will be performed by the program. The general goal of the scheduler is to dispatch jobs that satisfy the request, and to use the app versions that will run fastest.

The scheduler loops over processing resources, handling GPUs first so that they'll be kept busy if the disk space limit is reached (as it dispatches jobs, the scheduler keeps track of how much disk space is still available). For each processing resource *R*, the scheduler scans the job cache, starting at a random point to reduce lock conflict, and creates a **candidate list** of possible jobs to send, as follows.

For each scanned job *J*, we find the app versions that a) use a platform supported by *H*; b) use *R*; c) pass homogeneous redundancy and homogeneous app version constraints, and d) pass the plan class test. If no such app version exists, we continue to the next job. Otherwise we select the app version *V* for which *proj_flops(H, V)* is greatest.

We then compute a **score** for *J*, representing the "value" of sending it to *H*. The score is the weighted sum of several factors:

- If *J* has keywords (see section 2.4), and the volunteer has keyword preferences, whether *J*'s keywords match the volunteer's "yes" keywords; skip *J* if it has a "no" keyword.
- The allocation balance of the job submitter (see Section 3.9).
- Whether *J* was skipped in handling a previous scheduler request. The idea is that if *J* is hard to send, e.g. because of large RAM requirements, we should send it while we can.
- If the app uses locality scheduling, whether *H* already has the input files required by *J*.
- If the app has multi-size jobs, whether *proj_flops(H, V)* lies in the same quantile as the size of *J* (see section 3.5).

We then scan the candidate list in order of descending score. For each job *J*, we acquire a semaphore that protects access to the job cache, and check whether another scheduler process has already dispatched *J*, and skip it if so.

We then do a "fast check" (no database access) to see whether we can actually send *J* to *H*, given the jobs we have already decided to send. We skip *J* if a) the remaining disk space is insufficient; b) *J* probably won't complete by its deadline: i.e. if *queue_dur(R) + est_runtime(J, H, V) > delay_bound(J)*; or c) we are already sending another instance of the same job as *J*.

We then flag *J* as "taken" in the job cache, release the mutex, and do a "slow check" involving database access: Have we already sent an instance of *J* to this volunteer? Has *J* errored out since we first considered it? Does it still pass the homogeneous redundancy check? If so, skip *J* and release it in the cache.

If we get this far, we add the pair *(J, V)* to the list of jobs to send to client, and free the slot in the job cache. Letting *E* be the estimated scaled runtime of the job, we add *E* to *queue_dur(R)*, subtract *E* from *req_runtime(R)*, and subtract *J*'s usage of *R* from *req_idle(R)*. If both of the latter are nonpositive, we move on to the next processing resource. Otherwise we move to the next job in the candidate list.

## 7. Credit

BOINC grants **credit** – an estimate of FLOPs performed - for completed jobs. The total and exponentially-weighted recent average credit is maintained for each computer, volunteer, and team. Credit serves several purposes: it provides a measure of progress for individual volunteers; it is the basis of competition between volunteers and teams; it provides a measure of computational throughput for projects and for BOINC as a whole; and it can serve as proof-of-work for virtual currency systems.

To support these uses, the credit system ideally should:

- be **device neutral**: similar jobs (in particular the instances of a replicated job) should get about the same credit regardless of the host and computing resource on which they are processed;
- be **project neutral**: a given device should get about the same credit per time regardless of which project it computes for;
- resist **credit cheating**: i.e. efforts to get credit for jobs without actually processing them.

The basic formula is that one unit of credit represents one day of a CPU with a Whetstone benchmark of 1 GFLOPS. Credit is assigned in different ways depending on the properties of the application. If all jobs do the same computation, the project can time the job on a typical computer with known Whetstone benchmark, compute the credit accordingly, and tell BOINC to grant that amount per job. Similarly, if the app consists of a loop that executes an unpredictable number of times but that always does the same computation, the credit per iteration can be computed ahead of time, and credit can be granted (in the validator) based on the number of loop iterations performed.

For applications not having these properties, a different approach is needed. The BOINC client computes the peak FLOPS of processing resources as described earlier. The actual FLOPS of a particular application will be lower because of memory access, I/O, and synchronization. The **efficiency** of the application on a given computer is actual FLOPS divided by peak FLOPS. The efficiency of a CPU application can vary by a factor of 2 between computers; hence computing credit based on runtime and benchmarks would violate device neutrality. The problem is worse for GPUs, for which efficiencies are typically much less than for CPUs, and vary more widely.

To handle these situations, BOINC uses an adaptive credit system. For a completed job instance *J*, we define its **peak FLOP count** *PFC(J)* as:

$$PFC(J) = \sum_{r \in R} runtime(J) * usage(r) * peak\_flops(r)$$

where *R* is the set of processing resources used by the app version.

The server maintains, for each (host, app version) pair and each app version, the statistics of *PFC(J)/est_flop_count(J)* where *est_flop_count(J)* is the *a priori* job size estimate.

The **claimed credit** for *J* is *PFC(J)* times two normalization factors:

- Version normalization: the ratio of the version's average PFC to that of the version for which average PFC is lowest (i.e. the most efficient version).
- Host normalization: the ratio of the (host, app version)'s average PFC to the average PFC of the app version.

The instances of a replicated job may have different claimed credit. BOINC computes a weighted average of these, using a formula that reduces the impact of outliers, and grants this amount of credit to all the instances.

In the early days of BOINC we found that volunteers who had accumulated credit on one project were reluctant to add new projects, where they'd be starting with zero credit. To reduce this effect we introduced the idea of **cross-project credit**: the sum of a computer's, volunteer's, or team's credit over all projects in which they participate. This works as follows:

- Each host is assigned a unique **cross-project ID**. The host ID is propagated among the projects to which the host is attached, and a consensus algorithm assigns a single ID.
- Each volunteer is assigned a unique cross-project ID. Accounts with the same email address are equated. The ID is based on email address, but cannot be used to infer it.
- Projects export (as XML files) their total and recent average credit for hosts and volunteers, keyed by cross-project ID.
- **Cross-project statistics web sites** (developed by 3rd parties) periodically read the statistics data from all projects, collate it based on cross-project IDs, and display it in various forms (volunteer and host leaderboards, project totals, etc.).

## 8. Supporting volunteers

BOINC is intended to be usable for computer owners with little technical knowledge – people who may have never installed an application on a computer. Such people often have questions or require help. We've crowd-sourced technical support by creating systems where experienced volunteers can answer questions and solve problems for beginners. Many issues are handled via message boards on the BOINC web site. However, these are generally in English and don't support real-time interaction. To address these problems we created a system, based on Internet telephony, that connects beginners with questions to helpers who speak the same language and know about their type of computer. They can then interact via voice or chat.

BOINC volunteers are international – most countries are represented. All of the text in the BOINC interface – both client GUIs and web site – is translatable, and we crowd-source the translation. Currently 26 languages are supported.

## 9. Related work

Luis Sarmenta coined the term "volunteer computing" [20]. The earliest VC projects (1996-97) involved prime numbers (GIMPS) and cryptography (distributed.net). The first large-scale scientific projects were Folding@Home and SETI@home, which launched in 1999.

Several systems have been built that do volunteer computing in a web browser: volunteers open a particular page and a Java or Javascript program runs [21]. These systems haven't been widely used for various reasons: volunteers eventually close the window, C and FORTRAN science applications aren't supported, and so on.

In the context of BOINC, researchers have studied the availability of volunteer computers [5] [22] and their hardware and software attributes [23].

It is difficult to study BOINC's scheduling and validation mechanisms in the field. Initially, researchers used simulation to study client scheduling policies [24] [25], server policies [26], and entire VC systems [27]. As the mechanisms became more complex, it was difficult to model them accurately, and researchers began using emulation – simulators using the actual BOINC code to model client and server behavior (we restructured the code to make this possible). Estrada et al. developed a system called EmBoinc that combines a simulator of a large population of volunteer hosts (driven either by trace data or by a random model) with an emulator of a project server – that is, the actual server software modified to take input from the population simulator, and to use virtual time instead of real time [28]. Our group developed an emulator of the client [29]. Volunteers experience problems can upload their BOINC state files, and run simulations, through a web-based interface. This lets us debug host-specific issues without access to the host.

Much work has been done toward integrating BOINC with other HTC systems. Working with the HTCondor group, we developed an adaptor that migrates HTCondor jobs to a BOINC project. This is being used by CERN [30]. Other groups have built systems that dynamically assign jobs to different resources (local, Grid, BOINC) based on job size and load conditions [31]. Andrzejak et al. proposed using volunteer resources as part of clouds [32]. The integration of BOINC with the European Grid Infrastructure (EGEE) is described by Visegradi et al. [33] and Kacsuk et al. [34]. Bazinet and Cummings discuss strategies for subdividing long variable-sized jobs into jobs sized more appropriately to VC [35].

Research has been done on several of the problems underlying volunteer computing, such as validating results [36] [37] [38], job runtime estimation [39], accelerating the completion of batches and DAGs of jobs [40] [41], and optimizing replication policies [42] [43]. Many of these ideas could and perhaps will be implemented in BOINC. Cano et al. surveyed the security issues inherent in VC [44].

Files such as executables must be downloaded to all the computers attached to a project. This can impose a heavy load on the project's servers and outgoing Internet connection. To reduce this load, Elwaer et al. [45] proposed using a peer-to-peer file distribution system (BitTorrent) to distribute such files, and implemented this in a modified BOINC client. This method hasn't been used by any projects, but it may be useful in the future.

Researchers have explored how existing distributed computing paradigms and algorithms can be adapted to VC. Examples include evolutionary algorithms [46] [47], virtual drug discovery [48], MPI-style distributed parallel programming [49], Map-Reduce computing [50], data analytics [51], machine learning [52] [53] [54], gene sequence alignment [55], climate modeling [56], and distributed sensor networks [57]. Desell et al. describe a system that combines VC and participatory "citizen science" [58].

Silva et al. proposed extending BOINC to allow job submission by volunteers [59]. Marosi et al. proposed a cloud-like system based on BOINC with extensions developed at SZTAKI [15].

## 10. Conclusion and future work

We have described the features, architecture, and implementation of BOINC. The system has achieved the goal of transforming a large, heterogeneous, and untrusted pool of consumer devices into a reliable computing resource for a wide range of scientific applications. We now describe current and prospective future work aimed at extending the use of BOINC.

### 10.1 Coordinated volunteer computing

The original BOINC participation model was intended to result in a dynamic and growing ecosystem of projects and volunteers. This has not happened: the set of projects has been mostly static, and the volunteer population has gradually declined. The reasons are likely inherent in the model. Creating a BOINC project is risky: it's a significant investment, with no guarantee of any volunteers, and hence of any computing power. Publicizing VC in general is difficult because each project is a separate brand, presenting a diluted and confusing image to the public. Volunteers tend to stick with the same projects, so it's difficult for new projects to get volunteers.

To address these problems, we have created a new **coordinated model** for VC. This involves a new account manager, **Science United** (SU), in which volunteers register for scientific areas (using the

keyword mechanism described in Section 2.4) rather than for specific projects [60]. SU dynamically attaches hosts to projects based on these science preferences. Projects are vetted by SU, and SU has a mechanism (based on the linear-bounded model described above) for allocating computing power among projects. This means that a prospective new project can be guaranteed a certain amount of computing power before any investment is made, thus reducing the risk involved in creating a new project.

## 10.2  Integration with existing HTC systems

While the coordinated model reduces barriers to project creation, the barriers are still too high for most scientists. To address this, we are exploring the use of VC by existing HTC providers like computing centers and science gateways. Selected jobs in such systems will be handled by VC, thus increasing the capacity available to the HTC provider's clients, which might number thousands of scientists. We are prototyping this type of facility at Texas Advanced Computing Center (TACC) and at nanoHUB, a gateway for nanoscience [61]. Both projects use Docker and a universal VM app to handle application porting.

## 10.3  Data archival on volunteer devices

Host churn means that a file stored on a host may disappear forever at any moment. This makes it challenging to use BOINC for high-reliability data archival. However, it's possible. Encoding (e.g. Reed-Solomon) can be used to store a file in a way that allows it to be reconstructed even if a number of hosts disappear. This can be used to achieve arbitrarily high levels of reliability with reasonable space overhead. However, reconstructing the file requires uploading the remaining chunks to a server. This imposes high network overhead, and it requires that the original file fit on the server. To solve these problems, we have developed and implemented a technique called "multi-level encoding" in which the chunks from a top-level encoding are further encoded into smaller $2^{nd}$-level chunks. When a failure occurs, generally only a top-level chunk needs to be reconstructed, rather than the entire file.

## 10.4  Using more types of devices

Ideally, BOINC should be available for any consumer device that's networked and can run scientific jobs. This is not yet the case. BOINC runs on Android but not Apple iOS, because Apple's rules bar apps from dynamically downloading executable code. It also is not available for video game consoles or Internet of Things devices such as appliances. The barriers are generally not technical, but rather that the cooperation of the device manufacturers is needed.

In addition, although BOINC on Android detects on-chip GPUs via OpenCL, no projects currently have apps that use them. It's likely that their use will raise issues involving heat and power.

## 10.5  Secure computing

Recent processor architectures include features that allow programs to run in secure "enclaves" that cannot be accessed by the computer owner. Examples include Intel's Software Guard Extensions (SGX), AMD Secure Encrypted Virtualization, ARM's TrustZone, and the proposed Trusted Platform Module [62]. Extending BOINC to use such systems has two possible advantages: a) it could eliminate the need for result validation, and b) it would make it possible to run applications whose data is sensitive – for example, commercial drug discovery, and medical research applications involving patient data.

## 10.6 Recruiting and retaining volunteers

Early volunteer computing projects such as SETI@home received mass media coverage in 1999-2001, and attracted on the order of 1M volunteers.  Since then there has been little media coverage, and the user base has shrunk to 200K or so.  Surveys have shown that most volunteers are male, middle-aged, IT professionals [63] [64].

Several technology companies have promoted VC by developing and publicizing "branded" versions of the BOINC client.  Examples include Intel (Progress thru Processors), HTC (Power to Give) and Samsung (Power Sleep).  None of these resulted in a significant long-term increase in volunteers.

So volunteer computing faces the challenge of marketing itself to the broader public.  Possible approaches include social media, public service announcements, and partnerships with companies such as video game frameworks, device manufacturers, and cell phone service providers.

Another idea is to use BOINC-based computing as proof-of-work for virtual currency and other blockchain systems, and use the tokens issued by these systems as a reward for participating.  One such system exists [65], and others are under development.

## 10.7 Scheduling

There are a number of possible extensions and improvements related to scheduling; some of these are explored by Chernov et al. [66].  Dinis et al. proposed facilitating experimentation by making the server scheduler "pluggable" [67].

BOINC currently has no direct support for directed acyclic graphs (DAGs) of dependent jobs. BOINC doesn't currently support **low-latency computing**, by which we mean the ability to complete a job or DAG of jobs as fast as possible, and to give an estimate of completion time in advance.  Doing so would require a reworking of all the scheduling mechanisms.  For example, job runtime estimation would need to use sample variance as well as sample mean, in order to bound the probability of deadline miss.  Job deadlines would be assigned dynamically during job dispatch, rather than specified by the submitter.  Replication could be used to reduce expected latency.

When using host availability as part of job runtime estimation, BOINC uses the average availability.  This doesn't take into account the possibility that unavailability occurs on large chunks, or that it occurs with a daily or weekly periodicity.  Doing so could provide more accurate, and in some cases lower, estimates.

To accelerate the completion of batches (and perhaps DAGs), BOINC could add a mechanism to deal with the **straggler effect**: the tendency for a few jobs to delay the completion of the batch as a whole.

The idea is assign these jobs to fast, reliable, and available computers, and possibly to replicate the jobs. These ideas have been explored by others [40].

Binti et al. studied a job dispatch policy designed to minimize air-conditioning costs in campus grids [68].

SOMETHING ABOUT QOS

There are cases where BOINC has a negative impact on device performance. It may cause cooling fans to go on high speed, which disturbs some users. When a user is streaming a move, BOINC may decide that the device is idle and run GPU jobs, causing glitches in the movie playback. When BOINC starts up, it may start a number of large-memory jobs simultaneously, saturating disk I/O and slowing down the system. These issues can and should be fixed.

### 10.8 Use of cloud storage

In the current BOINC architecture, input and output files reside on storage that is local or NFS-attached to the server. This requires projects with large data to buy and maintain storage systems, and it may cause a network bottleneck at the server. In some cases projects could reduce cost and increase performance by using commercial or institutional cloud storage, accessed via RESTful APIs [69].

## Acknowledgements

Many people contributed to the ideas presented here and to their implementation. Thanks in particular to Bruce Allen, Nicolás Alvarez, Matt Blumberg, Karl Chen, Carl Christensen, Charlie Fenton, David Gedye, Francois Grey, Eric Korpela, Janus Kristensen, John McLeod VII, Kevin Reed, Ben Segal, Michela Taufer, and Rom Walton. The National Science Foundation has supported BOINC with several grants, most recently award #1664190.